\documentclass[aps,preprint,prl,superscriptaddress]{revtex4}
\usepackage{graphicx}

\begin{document}

\title{Robust formation of morphogen gradients}

\author{T. Bollenbach} 
\affiliation{Max-Planck-Institute for Physics of Complex Systems, N\"othnitzer
Str.~38, 01187 Dresden, Germany} 
\author{K. Kruse} 
\affiliation{Max-Planck-Institute for Physics of Complex Systems, N\"othnitzer
Str.~38, 01187 Dresden, Germany} 
\author{P. Pantazis} 
\affiliation{Max-Planck-Institute for Molecular Cell Biology and Genetics, Pfotenhauer
Str.~108, 01307 Dresden, Germany} 
\author{M. Gonz\'alez-Gait\'an} 
\affiliation{Max-Planck-Institute for Molecular Cell Biology and Genetics, Pfotenhauer
Str.~108, 01307 Dresden, Germany} 
\author{F. J\"ulicher}
\affiliation{Max-Planck-Institute for Physics of Complex Systems, N\"othnitzer
Str.~38, 01187 Dresden, Germany}

\date{\today}

\begin{abstract}
We discuss the formation of graded morphogen profiles in a
cell layer by
nonlinear transport phenomena, important for patterning developing
organisms.
We focus on a process termed transcytosis, where morphogen transport results from binding
of ligands to receptors on the cell surface, incorporation
into the cell and subsequent externalization. Starting from
a microscopic model, we derive effective transport
equations. We show that, in contrast to morphogen transport
by extracellular diffusion, transcytosis leads to robust
ligand profiles which are insensitive to the rate of ligand
production.
\end{abstract}

\maketitle

An essential feature during the development of an organism is 
the emergence of different cell types. This process, called cell
differentiation, is intimately linked to the position of the cells in
the organism. About fifty years ago, Turing suggested the existence of 
molecules that provide positional information
for differentiating cells \cite{turing1952}. Turing proposed that these morphogens
self-organize into patterns which structure the future organism.
Pattern formation based on reaction diffusion systems has been 
studied extensively~\cite{cross1993,koch1994,gierer1972}.
Today, a number of proteins have been identified to act as morphogens. 
In contrast to Turing's original suggestion, morphogens
are secreted from a spatially localized source and form a stable, 
long-ranged concentration gradient in the adjacent tissue \cite{wolpert1969}.
There, cells detect the morphogen by  receptors
located on the cell surface and respond with patterns of 
gene expression that depend on the morphogen concentration. 
In this way, cell fate depends on the distance from the morphogen 
source and the interplay of several morphogen gradients originating
from spatially distinct sources leads to the complex patterning of
developing organisms. 
For this to be possible, the gene expression patterns must be formed with
high precision and also need to be robust, i.e. they should not
depend sensitively on parameters that are likely to fluctuate. 
Robust gene expression can result from the robustness of 
the morphogen concentration gradient \cite{eldar2003,eldar2002} or can
be achieved by other mechanisms \cite{houchmandzadeh2002}. 
Interestingly, different species use very similar
morphogens and a given species can use the same 
morphogen at different stages during development. 
An example are morphogens 
of the TGF-$\beta$ super-family which exist
in a wide range of organisms ranging from the fruit fly 
Drosophila to humans.

The formation of morphogen 
gradients is still poorly understood. In particular, it is not clear, how 
morphogens are transported in the tissue adjacent to the source.
Several possibilities have been proposed. Secreted
morphogen molecules might simply diffuse in the extracellular space
between cells  \cite{crick1970} and for some morphogens this indeed seems to be
the dominant form of transport~\cite{mcdo97}.  
However, there is growing
experimental evidence against simple diffusion for several morphogens
\cite{entchev2000,bele04}. Other transport mechanisms, that have been
proposed, include the bucket brigade, where molecules are
"handed over" from one cell to adjacent cells \cite{kerszberg1998}, or 
transport by cell displacements in the tissue~\cite{pfei00}. 

\begin{figure}[htbp]
\includegraphics[scale=0.3]{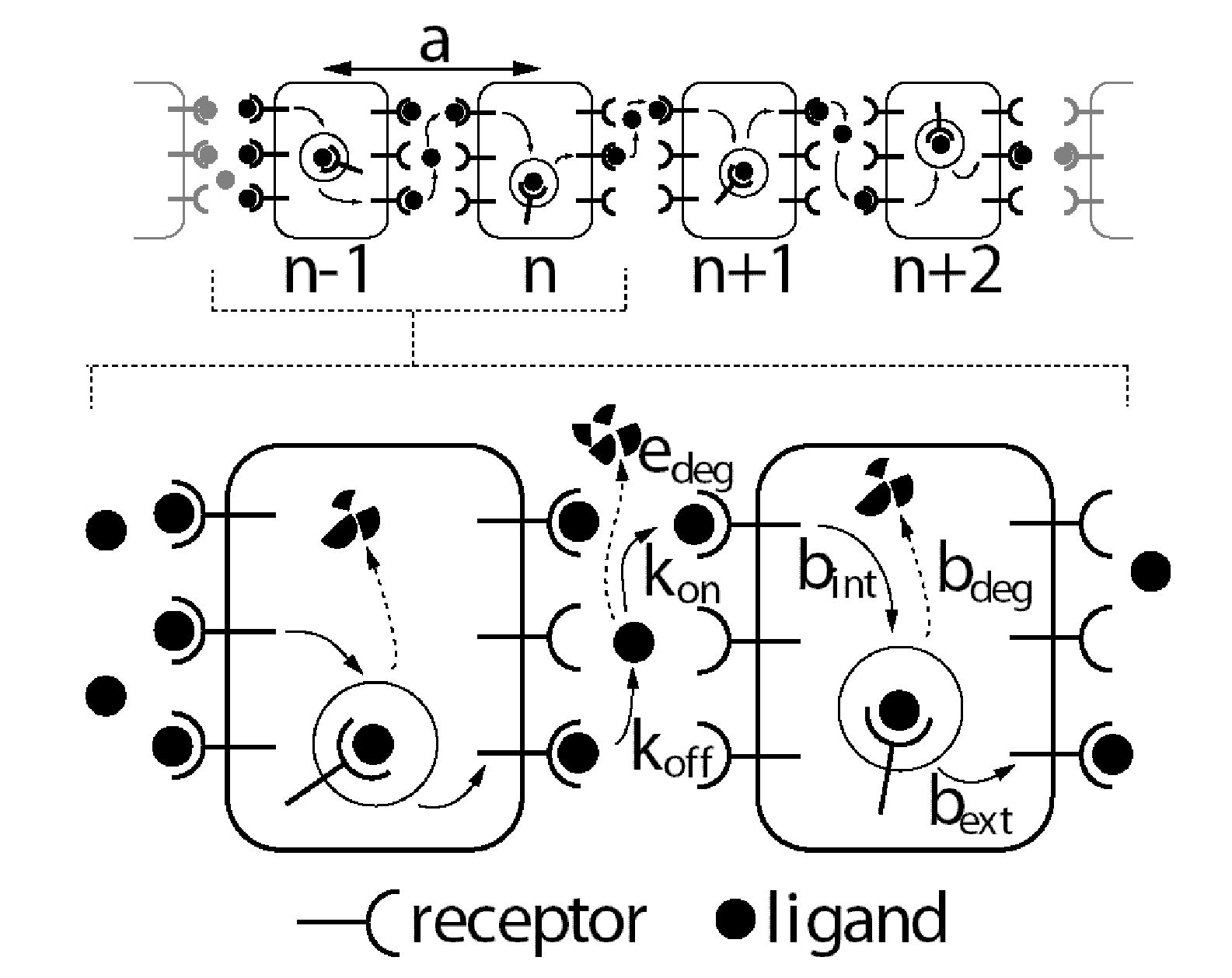}
\caption{\label{fig:discrete_model} Schematic representation of transport by
  transcytosis in a chain of cells of diameter $a$ indexed by $n$. The rates
  of ligand-receptor binding and unbinding, internalization and externalization
  of ligand-receptor pairs are denoted $k_{\rm on}$, $k_{\rm off}$, $b_{\rm int}$
  and $b_{\rm ext}$.  Degradation of ligand occurs inside the cells with rate $b_{\rm deg}$
  and in the extracellular space with rate $e_{\rm deg}$.}
\end{figure}

Another possibility is suggested by recent experiments on the
morphogen Dpp \cite{entchev2000}. Dpp belongs to the 
TGF-$\beta$ super family and
patterns  the developing wing discs of Drosophila. The wing disc is a 
sheet formed from one layer of cells
that is the precursor of a wing.
Fluorescently labeled Dpp was tracked in
wing discs where it is secreted by a narrow strip of cells from where
it spreads into the adjacent tissue.
If all cells adjacent to the secreting cells are defective in the internalization of the 
morphogen
into the cell, 
Dpp gradients extend only  about two cells from the source as 
compared to 30 cells in a non-mutant fly. If only cells in a small patch close to 
the source show this defect, a pronounced transient  depletion of the 
morphogen is observed behind the mutant patch during gradient formation.
This suggests that Dpp is not simply diffusing through the
extracellular space. 
It has been suggested that these observations can be described by a
scenario where Dpp diffuses in the extracellular space, binds to and is released from
receptors \cite{lander2002}. 
However, an analysis which takes into account more
recent experimental results  indicates that free diffusion alone cannot
account for the observations \cite{kruse2004}.
An alternative suggestion is that 
long range transport of Dpp is generated by transcytosis, i.e. 
by repeated rounds of morphogen binding to cell
surface receptors, internalization 
into the cell and subsequent externalization and release of 
the ligand from the receptor at a different point on the cell surface \cite{entchev2002}.
A simple model to describe transcytosis is
illustrated in Fig.~\ref{fig:discrete_model}. Ligands spread from a stripe
of secreting cells into a two-dimensional tissue. However, in many situations 
a one-dimensional description
of ligand profiles as a function of the distance $x$ to the source
is appropriate \cite{lander2002,kruse2004}.
Transcytosis, can be characterized by rates $k_{\rm on}$ and 
$k_{\rm off}$ for receptor binding
and unbinding of free ligands, as well as
$b_{\rm int}$ and $b_{\rm ext}$ for internalization and externalization
of ligand-bound receptors. Internalized 
morphogens are degraded with rate $b_{\rm deg}$. 
The free diffusion of ligands in the narrow space between cells is characterized
by the diffusion coefficient $D_0$.
Diffusion alone does not lead to efficient long-range transport if the diffusion 
length $\xi_d=(D_0/e_{\rm deg})^{1/2}$ is smaller than one 
or several cell diameters. Here, $e_{\rm deg}$ is the extracellular degradation rate.

\begin{figure}[htbp]
\includegraphics[scale=0.2]{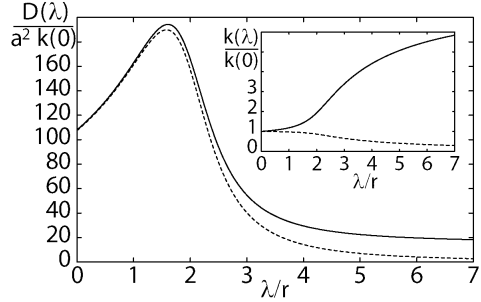}
\caption{\label{fig:diffusion_coefficient} Effective diffusion coefficient $D(\lambda)$
as a function of ligand concentration $\lambda$ for 
  $D_0/a^2 b_{\rm deg}=10$ (solid line) and
   $D_0=0$ (dashed line).
Inset: effective degradation rate $k(\lambda)$ as
  a function of $\lambda$ for $e_{\rm deg}=0$ (dashed
  line) and $e_{\rm deg}/b_{\rm deg}=5$ (solid line).
Parameters are: $b_{\rm int}/b_{\rm deg}=b_{\rm
  ext}/b_{\rm deg}=3\times 10^3$, $k_{\rm on}R/b_{\rm deg}=1.1\times 10^4$, $k_{\rm
  off}/b_{\rm deg}=7\times 10^2$. 
 }
\end{figure}

On scales large compared to the cell diameter $a$, this model
together with the assumption that receptors are also synthesized and degraded in the cells,
leads to effective
transport equations for the number densities $\lambda(t,x)$ of morphogens and
$\rho(t,x)$ of receptors
given by 
\begin{eqnarray}
\label{eq:dldt}
\partial_t \lambda & = & \partial_x(D(\lambda,\rho)\partial_x\lambda-
D_\rho(\lambda,\rho)\partial_x\rho)-
k(\lambda,\rho)\lambda\\
\label{eq:drdt}
\partial_t\rho & = & \nu_{\rm syn}(\lambda,\rho) - \nu_{\rm deg}(\lambda,\rho)\rho
\quad.
\end{eqnarray}
Here, $D(\lambda, \rho)$ is an effective
diffusion coefficient and $k(\lambda,\rho)$ an effective
degradation rate that depend on both the ligand and receptor concentrations. 
Furthermore, a 
term exists 
which describes ligand currents induced by gradients of the receptor concentration.
We assume for simplicity that cells are non-polar such that externalization 
occurs with equal probability at any place on the cell surface.
As a consequence, the resulting transport is non-directional. 
The kinetics of the receptor density is characterized by effective source and
sink terms $\nu_{\rm syn}(\lambda,\rho)$ and $\nu_{\rm deg}(\lambda,\rho)$.
The dependences of the effective
diffusion coefficient $D$, the degradation rate $k$, the
coefficients $D_\rho$, $\nu_{\rm syn}$ and $\nu_{\rm deg}$ 
on the ligand and receptor concentrations
can be calculated explicitly \cite{bollenbach_unpublished}.
In order to keep our discussion simple,
we focus here on the case where the surface receptor number 
per cell
$R$
is maintained constant. Then, ligand transport is 
described by a single nonlinear diffusion equation
\begin{equation}
\label{eq:dldt_rconst}
\partial_t\lambda=\partial_x(D(\lambda)\partial_x\lambda)-k(\lambda)\lambda\quad.
\end{equation}
Explicit expressions of $D(\lambda)$ and $k(\lambda)$ 
are given in \footnote{We find $k(\lambda) = C_{+}(\lambda) \left[b_{\rm deg}{b_{\rm int}}/[2
    {b_{\rm ext}}\left( {b_{\rm ext}} + {b_{\rm int}} \right)
    ]
+ {e_{\rm deg}}{k_{\rm off}}/
    C_{-}(\lambda) \right] /\kappa\lambda$. For $D_{0}=0$,  we furthermore have
$D(\lambda) =  [a^2{b_{\rm ext}}{b_{\rm int}}{k_{\rm off}}\kappa r 
   C_{-}(\lambda) ]/[4 A(\lambda)
    [ 2\kappa r k_{\rm off}({b_{\rm ext}}+ 
         {b_{\rm int}})  + 
      {b_{\rm int}}C_{-}(\lambda)  ] ]$. In these expressions  
      $\kappa=ak_{\rm on}$,
$A(\lambda) =  \left[-4 b_{\rm ext} (b_{\rm ext}+b_{\rm int})\kappa^2 r\lambda+(
  b_{\rm int} \kappa r + b_{\rm ext} B_{+}(\lambda))^2\right]^{1/2}$, 
  $B_{\pm}(\lambda)={k_{\rm off}} + \kappa(\lambda\pm r )$, $C_{\pm}(\lambda)=b_{\rm int}\kappa r\mp A(\lambda)\pm b_{\rm ext}B_{\pm}$.}.
The coefficients are 
displayed as functions of the ligand concentration $\lambda$
in Fig. ~\ref{fig:diffusion_coefficient}. 
For small $\lambda$, $D(\lambda)$ and $k(\lambda)$ 
approach finite values, and the transport
equation becomes linear. 
For large $\lambda$, $D(\lambda) \approx D_0 + c_1/\lambda^2$ with
$c_1=a b_{\rm int} k_{\rm off} r/(4k_{\rm on})$, where $r=R/a$. 
The effective diffusion coefficient $D(\lambda)$ decreases for large ligand concentration 
because receptors are increasingly occupied and not available for transport.
A maximum of $D$
occurs for intermediate values of $\lambda$, if the ligand binding rate $k_{\rm on}$
exceeds a critical value.
This implies that for small $\lambda$, $D$ increases for increasing $\lambda$.
Similarly, the effective degradation rate exhibits for large
$\lambda$ the asymptotic behavior 
$k\approx e_{\rm deg} + c_2/\lambda$ where 
$c_2=b_{\rm deg}b_{\rm int}r/b_{\rm ext}$. 

\begin{figure}[htbp]
\includegraphics[scale=0.2]{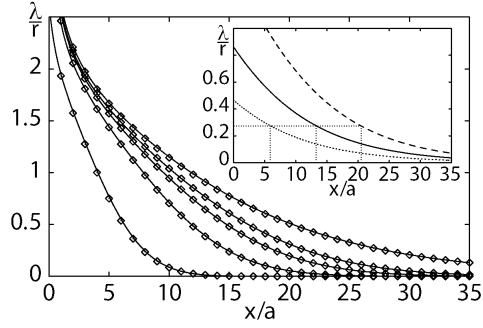}
\caption{\label{fig:time_dev}
Ligand densities $\lambda(x)$ in the presence of a source at
$x=0$ at different times $b_{\rm deg} t=0.18, 0.54, 0.9, 1.26$ during gradient
formation and in the steady state. Lines indicate solutions to Eq. (\ref{eq:dldt_rconst}),
while symbols indicate solutions to Eqns. (\ref{eq:freeligand})-(\ref{eq:internalbound})
for comparison.
The robustness of the steady state profile
is ${\mathcal{R}}\approx 470$. 
Parameters as in Fig.~\ref{fig:diffusion_coefficient} with $D_0=0$, $j_{0}/b_{\rm deg}R=70$
and $j=0$ at $x/a=50$.
Inset: steady state ligand profiles 
for same parameters but  $j_0/b_{\rm deg}R=7$ where 
${\mathcal{R}}\approx 0.1$. The profile (solid line) is strongly affected 
by halving (dotted line) or doubling (dashed line) 
the ligand current of the reference state.
}
\end{figure}

We solve Eq. (\ref{eq:dldt_rconst}) for $x>0$, representing the space adjacent to
the ligand source, and with boundary conditions $j(x=0)=j_0$ and $j=0$ for large $x$.
Here, $j_0$ is proportional to the rate 
of ligand secretion
in the source.
In this situation,
a graded ligand profile is generated and maintained in the steady state, see Fig.~\ref{fig:time_dev}.

The effect of the source is captured by the boundary condition on the current $j=j_0$ at $x=0$, where 
$j=-D(\lambda) \partial_{x}\lambda$, see Eq.~(\ref{eq:dldt_rconst}). If initially
at time $t=0$
ligand is absent
 $\lambda=0$,
the ligand spreads 
for $t>0$
into the region $x\ge0$, and builds up a 
gradient. In the steady state, the total ligand concentration decreases 
monotonically with increasing distance from the source. 
The steady state satisfies $\partial_x j+k(\lambda)\lambda =0$ and 
is reached on a time scale of the order 
of $1/k(\lambda=0)$. 
 The corresponding ligand distribution $\lambda(x)$
obeys
\begin{equation}
x=-\int_{\lambda(0)}^{\lambda (x)}d\lambda^{\prime} \;
D(\lambda^{\prime})/j(\lambda^{\prime}) \label{eq:stst}
\quad,
\end{equation}
where the ligand current $j$ in the steady state is given by
\begin{equation}
j (\lambda) = \left [2\int_0^{\lambda}d\lambda^{\prime}\ k(\lambda^{\prime})
  D(\lambda^{\prime}) \lambda^{\prime}\right ]^{1/2}\quad. \label{eq:stst_current}
\end{equation}
For small $\lambda$, the ligand profile decays as 
$\lambda \propto \exp(-x/\xi)$ with 
$\xi=(D(0)/k(0))^{1/2}$. 
In the case $D_{0}=0$, the steady state profile has a singularity at
$x=x^{*}$:
\begin{equation}
\lambda\sim (x-x^*)^{-1}(-\ln{(x-x^*)})^{-1/2}\label{eq:sing}
\end{equation} 
Choosing $j_{0}>0$ ensures 
$x^*<0$ such that $x^{*}$ is not physically accessible.
This singularity results from
the asymptotic behavior of the current $j$ for large $\lambda\gg \lambda_T$:
$j(\lambda)^2\approx j(\lambda_T)^2+2
  e_{\rm deg} c_1 \ln(\lambda/\lambda_T) + 2 c_1 c_2
  (1/\lambda_T-1/\lambda)$. 
Here,
$\lambda_T$ denotes a characteristic ligand concentration beyond which the 
limit of large ligand concentration holds. 
Therefore the current
  diverges as $j^2 \approx 2 c_1 e_{\rm
  deg}\ln \lambda$. Note, that for $e_{\rm deg}=0$ the current 
  reaches for large $\lambda$ a finite
 maximal value $j_{\rm max}$ and the corresponding steady state ligand 
 profile diverges as $\lambda \approx c_1/[j_{\rm max} (x-x^*)]$.
  
The singular behavior of the steady state profile near $x=x^*$ has remarkable
consequences for the robustness of gradient formation. 
Such a robustness can
be quantified by an appropriate response function of the system \cite{eldar2003}.
We define the robustness ${\mathcal{R}}(j_0):=a (j_0 \partial_{j_0}x(\lambda))^{-1}$, 
where the function $x(\lambda)$ is given by Eq.~(\ref{eq:stst}).   
It becomes large if changes of $j_0$ have little influence on the
ligand profile in the steady state. 
Here, a robustness of 
${\mathcal{R}}=1$ implies that under a 100\% increase
of $j_0$ the position at which the ligand profile attains any fixed value
is displaced by about one cell diameter $a$. Thus for ${\mathcal{R}}\geq1$, cells in the target tissue
cannot detect variations of the ligand concentration
due to significant changes of $j_0$,
 while for ${\mathcal{R}}\leq1$ 
the steady state profile is strongly affected, see Fig.~\ref{fig:time_dev}.
Note, that 
${\mathcal{R}}$ is independent of $\lambda$.
Indeed, ${\mathcal{R}}$ can be rewritten as
\begin{equation}
{\mathcal{R}} = a (j_0 \partial_{j_0}x)^{-1} =  a
\partial_{\lambda_0}j_0/D(\lambda_0) = a k(\lambda_0) \lambda_0/j_0
\end{equation}
where $\lambda_0=\lambda(x=0)$ and Eqs.~(\ref{eq:stst}) and 
(\ref{eq:stst_current}) have been
used. The robustness is thus 
completely determined
by the ratio of the effective degradation rate and the ligand current 
at $x=0$. High
degradation rates and small currents lead to a robust gradient. Using the asymptotic behavior of the steady state profile for $D_0=0$, 
we find that the robustness increases rapidly for large currents $j_0$
as ${\mathcal{R}}\sim j_0^{-1} e^{j^2_0/j^2_c}$ with $j^2_c = 1/2 c_1 e_{\rm deg} $.
For small $j_0$, ${\mathcal{R}}\simeq a/\xi$ becomes constant.

\begin{figure}[htbp]
\includegraphics[scale=0.2]{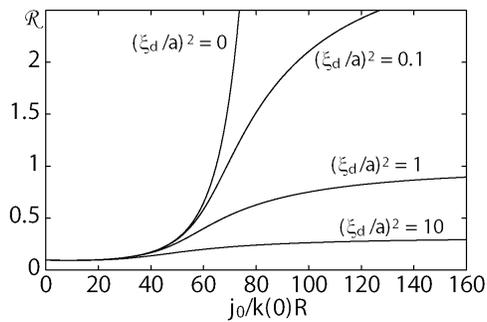}
\caption{\label{fig:robustness}Robustness ${\mathcal{R}}$ of steady state ligand
profiles as a
function of the ligand current $j_{0}$ from the source for different values of the ratio 
of the diffusion length $\xi_{d}$ and the cell size $a$. 
Parameters as in Fig.~\ref{fig:diffusion_coefficient}.
}
\end{figure}

The situation is different if $D_0$ is finite. In this case, both $k(\lambda)$ and $D(\lambda)$
are nonvanishing and constant for large $\lambda$ and consequently
Eq.~(\ref{eq:dldt_rconst})
becomes linear in this limit.
The  steady state profile for $\lambda\gg c_1/D_0$ behaves as
$\lambda\sim \exp(-x/\xi_d)$. In the interval $\lambda_T<\lambda<c_1/D_0$
the ligand profile is well described by Eq.~(\ref{eq:sing}), 
while for small $\lambda\ll \lambda_T$,
$\lambda\sim \exp(-x/\xi)$. As a consequence of finite $D_0$, 
the robustness approaches a finite value
$\mathcal{R}_{\rm max} = a/\xi_d$ for large $j_0$.
For $\xi_{d}<a$, this implies
${\mathcal{R}}>1$ for large $j_0$ and cells 
in the target tissue are insensitive to variations of $j_0$ by a factor of two.
The robustness is displayed 
in Fig.~\ref{fig:robustness} as a function of $j_0$ for different values of $\xi_{d}/a$. 

Finally, we briefly discuss how the transport Eq. (\ref{eq:dldt_rconst})
is derived from a microscopic model.  We denote the
number of free ligands in the space between cell $n$ and $n+1$ by $L_n$ and
the numbers of receptor bound ligands inside cell $n$ by $S^{(i)}_n$.
Similarly, we introduce the numbers $S_n^{(r)}$ and $S_n^{(l)}$ of receptor-bound
ligand on the right and left surface of cell $n$, respectively, see Fig.~\ref{fig:discrete_model}.
The dynamics of these quantities is given by
\begin{eqnarray}
\partial_t L_n & = & k_{\rm off} (S_n^{ (r)} + S_{ n+1}^{ (l)}) - k_{\rm on}
(R - S_n^{ (r)} - S_{ n+1}^{ (l)} ) L_n -e_{\rm deg} L_n\label{eq:freeligand} \\
\partial_t S_n^{ (r)} & = & -k_{\rm off} S_n^{ (r)} + k_{\rm on}
\Bigl(\frac{R}{2}-S_n^{ (r)}\Bigr) L_{ n} - b_{\rm int} S_{ n}^{ (r)} + \frac{1}{2} b_{\rm ext} S_n^{ (i)} \label{eq:rightsurface}\\
\partial_t S_n^{ (l)} & = & -k_{\rm off} S_n^{ (l)} + k_{\rm on}
\Bigl(\frac{R}{2}-S_n^{ (l)}\Bigr) L_{ n-1} - b_{\rm int} S_{ n}^{ (l)} + \frac{1}{2} b_{\rm ext} S_n^{ (i)} \label{eq:leftsurface}\\
\partial_t S_n^{ (i)} & = & - b_{\rm ext} S_n^{ (i)} + b_{\rm int} (S_{ n}^{ (l)} + S_n^{ (r)}) - b_{\rm deg} S_n^{ (i)} \label{eq:internalbound}.
\end{eqnarray}
On large scales, we describe the ligand profiles by the
densities  $l(x)=L_n/a$, $s(x)=S_n^{(i)}/a$ and 
$s_\pm(x)=[S_n^{ (l)}\pm S_n^{ (r)}]/a$, where $x=na$.
Because transport on large scales $L$ is governed by time scales
$\tau_L\simeq L^2/D$ which are long compared to the relaxation time
on the cellular scale $\tau_a\simeq a^2/D$, three of the dynamic 
equations can be adiabatically eliminated. 
The effective Eq. (\ref{eq:dldt_rconst}) for the total ligand
density $\lambda=l+s+s_+$ follows by taking a continuum limit 
\cite{bollenbach_unpublished}. A comparison of the full solutions of 
Eqs.~(\ref{eq:freeligand})-(\ref{eq:internalbound}) to Eq.~(\ref{eq:dldt_rconst})
is shown in Fig.~\ref{fig:time_dev}, demonstrating that the effective transport
Eq.~(\ref{eq:dldt_rconst}) captures the behavior of the microscopic model.

In conclusion, we have developed a general theoretical framework for
ligand transport and the generation of characteristic ligand profiles via transcytosis. 
Cell differentiation is triggered
at threshold levels of morphogen concentration.
The robustness of patterning has been demonstrated experimentally under
conditions where Dpp was overexpressed \cite{mori96,entchev2000,kruse2004}.
Reliable patterning can be achieved by robustly generating morphogen profiles.
We have
defined the robustness of steady state ligand profiles with respect to the rate of
ligand secretion $j_0$. We find that transcytosis naturally leads to a large
robustness of the profile. 
Free extracellular diffusion reduces 
robustness significantly if the diffusion length  $\xi_d$ exceeds
several cell sizes. 
The origin of robustness in transcytosis are nonlinearities in the 
ligand current.
A similar mechanism based on nonlinear degradation alone 
permits the robust formation of gradients in a free
diffusion model \cite{eldar2003}. Robustness
can also be achieved by other nonlinear effects \cite{eldar2002}.

In the general case where both ligand and receptor densities are taken into
account, new phenomena appear~\cite{bollenbach_unpublished}. In 
particular ligand currents can be driven by
receptor gradients. Furthermore, ligand gradients are accompanied by a gradient in the 
receptor density as is indeed observed experimentally \cite{entchev2000,teleman2000}.
In addition, if cells posess a polarity and define a direction in the tissue, transcytosis can 
lead to directed transport.
Finally, it is straightforward to generalize the concepts presented here to higher dimensions.

The effects discussed here become important if the diffusion length $\xi_d$ 
is smaller than the range $\xi\simeq 30 a$ of the gradient. 
From experiments where endocytosis has been blocked in the target
tissue, we estimate $\xi_d\leq 2a$ \cite{entchev2000}. 
The rates of transcytosis have to be sufficiently large to guarantee gradient
formation within the experimentally observed times of about $k(0)^{-1}\simeq $8 hours.
Assuming that  the rates $k_{\rm on}R$, $k_{\rm off} $, $b_{\rm int} $, and $b_{\rm ext} $
are of the order of $1/$min, we estimate $D\sim a^2/$min. This diffusion coefficient
is sufficient to generate a gradient over a length $\xi\simeq 30a$ during 8 hours.
Numerical solutions to the transcytosis model (see Fig. 3) confirm this
estimate. A rate of one per minute is larger than values measured in different but related systems 
\cite{lauf93} but is plausible for intracellular trafficking.

The values of the rate constants are difficult to measure while the effective diffusion
coefficient and the effective degradation rates are more easily accessible in
fluorescence recovery experiments. 
Our work suggests that
if transcytosis dominates transport, 
morphogen gradients are robust with respect to morphogen overexpression. This
can be directly tested in future experiments.

\end{document}